\documentclass[sigconf]{acmart}

\usepackage{multirow}
\usepackage[ruled,vlined,lined,commentsnumbered]{algorithm2e}
\usepackage{colortbl}
\usepackage{enumitem}
\usepackage{makecell}

\copyrightyear{2026}
\acmYear{2026}
\setcopyright{cc}
\setcctype{by}
\acmConference[SEAMS '26]{21st International Conference on Software Engineering for Adaptive and Self-Managing Systems}{April 13--14, 2026}{Rio de Janeiro, Brazil}
\acmBooktitle{21st International Conference on Software Engineering for Adaptive and Self-Managing Systems (SEAMS '26), April 13--14, 2026, Rio de Janeiro, Brazil}
\acmPrice{}
\acmDOI{10.1145/3788550.3794884}
\acmISBN{979-8-4007-2445-9/2026/04}

\begin{document}


\author{Khouloud Gaaloul}
\affiliation{%
  \institution{University of Michigan--Dearborn}
  \city{Dearborn}
  \state{MI}
  \country{USA}
}
\email{kgaaloul@umich.edu}

\author{Zaid Ghazal}
\affiliation{%
  \institution{University of Michigan--Dearborn}
  \city{Dearborn}
  \state{MI}
  \country{USA}
}
\email{zghazal@umich.edu}

\author{Madhu Latha Pulimi}
\affiliation{%
  \institution{University of Michigan--Dearborn}
  \city{Dearborn}
  \state{MI}
  \country{USA}
}
\email{pulimiml@umich.edu}

\author{Sam Emmanuel Kathiravan}
\affiliation{%
  \institution{University of Michigan--Dearborn}
  \city{Dearborn}
  \state{MI}
  \country{USA}
}
\email{samanuel@umich.edu}

\renewcommand{\shortauthors}{Gaaloul et al.}
\settopmatter{authorsperrow=4}


\title{Grammar-Constrained Refinement of Safety Operational Rules Using Language in the Loop: What Could Go Wrong}

\begin{abstract}
Safety specifications in cyber-physical systems (CPS) capture the operational conditions the system must satisfy to operate safely within its intended environment. As operating environments evolve, operational rules must be continuously refined to preserve consistency with observed system behavior during simulation-based verification and validation. Revising inconsistent rules is challenging because the changes must remain syntactically correct under a domain-specific grammar. Language-in-the-loop refinement further raises safety concerns beyond syntactic violations, as it can produce semantically unjustified refinements that overfit to the observed outcomes. We introduce a framework that combines counterfactual reasoning with a grammar-constrained refinement loop to refine operational rules, aligning them with the observed system behavior. Applied to an autonomous driving control system, our approach successfully resolved the inconsistencies in an operational rule inferred by a conventional baseline while remaining grammar compliant. An empirical large language model (LLM) study further revealed model-dependent refinement quality and safety lessons, which motivate rigorous grammar enforcement, stronger semantic validation, and broader evaluation in future work.
\end{abstract}




\maketitle

\section{Introduction}
\label{sec:intro}
Safety specifications for cyber-physical systems (CPS) rely on operational rules that translate safety requirements into testable conditions on system executions for verification and validation. In autonomous driving, these rules are scoped by the Operational Design Domain (ODD)~\cite{fraade2018measuring,garcia2022vehicle} and informed by ISO 26262~\cite{debouk2019overview} and ISO/PAS 21448 (SOTIF)~\cite{kirovskii2019driver}. They are typically expressed as logical or quantitative constraints over environmental and system variables, and evaluated through simulation-based testing~\cite{tuncali2018simulation} to ensure safe operation within the declared operational boundaries. However, as the operating conditions evolve, the observed system outcomes may shift and expose inconsistencies with the verdict of the operational rule. Maintaining rule validity and ensuring continued alignment with observed executions is a recurring challenge. This motivates an automated rule refinement process that restores consistency while preserving the underlying safety requirement and keeping the stated ODD unchanged.

Mining operational rules from observed traces has commonly been performed by learning linear temporal logic (LTL) properties~\cite{rescher2012temporal,pnueli1977temporal} and by applying specification mining and machine learning (ML) methods~\cite{jha2019telex,lemieux2015general}, typically constrained by templates rather than a domain-specific grammar. Other works use surrogate models~\cite{jodat2025automated,jodat2024test,zhu2022fuzzy} and genetic programming (GP)~\cite{gaaloul2021combining} as interpretable methods to infer and correct STL properties, or apply parameter mining and falsification~\cite{asarin2011parametric,hoxha2018mining,kyriakis2019specification} to tune bounds that make observed behaviors satisfy a fixed specification. No prior work is designed to automatically refine an existing inconsistent operational rule under a domain-specific grammar. Recent work has explored using large language models (LLMs) to support the construction and maintenance of safety specification artifacts. Nouri et al.~\cite{nouri2024engineering} propose a prompt-based pipeline for an automotive SafetyOps workflow that generates safety requirements and then checks the resulting rule set for redundancy, contradictions, and other quality issues. Li et al.~\cite{li2025automatic} use LLMs to generate LTL specifications under safety restrictions, then iteratively refine candidate formulas using language inclusion checks and counterexamples. Both works commonly combine LLM generation with automated validation. However, to our knowledge, no prior work uses grammar-constrained LLMs to refine an existing inconsistent operational rule to align with observed system behavior. Moreover, validation in LLM-based specification synthesis pipelines typically targets syntactic or format compliance, rather than consistency with runtime outcomes. We propose an approach for refining operational rules to resolve inconsistencies between rule verdicts and observed system behavior. Refinement is restricted to operational rules that operationalize a fixed safety requirement within a stated ODD. Rather than constructing a formal proof of requirement satisfaction, our approach is guided by observed counterfactual evidence. The core novelty is combining counterfactual analysis with a grammar-constrained LLM guided by a domain-specific grammar to synthesize minimal refinements of operational rules while preserving syntactic correctness and semantic validity. The main contribution of this paper is a rule refinement framework that combines counterfactual reasoning to localize inconsistency boundaries between operational rules and observed behavior, and grammar-constrained LLMs to generate interpretable, syntactically valid refinements to the operational rules that capture a pre-specified safety requirement within a stated ODD. We report an initial empirical study on an autonomous driving subsystem (ADS), evaluating effectiveness against a conventional baseline. We further assess the model-dependent language-in-the-loop refinement quality across multiple LLM variants and derive safety lessons.

\textbf{Organization.} The remainder of this paper is organized as follows. Section~\ref{sec:approach} presents the operational rule refinement framework, including the rule semantics, the domain-specific grammar, and the counterfactual guided, grammar constrained refinement loop. Section~\ref{sec:eval} reports our experimental exploration on an autonomous driving subsystem and summarizes the results of the LLM variant study. Section~\ref{sec:conclusion} concludes and outlines directions for future work.

\begin{figure}[tb]
    \centering
    \includegraphics[width=\linewidth]{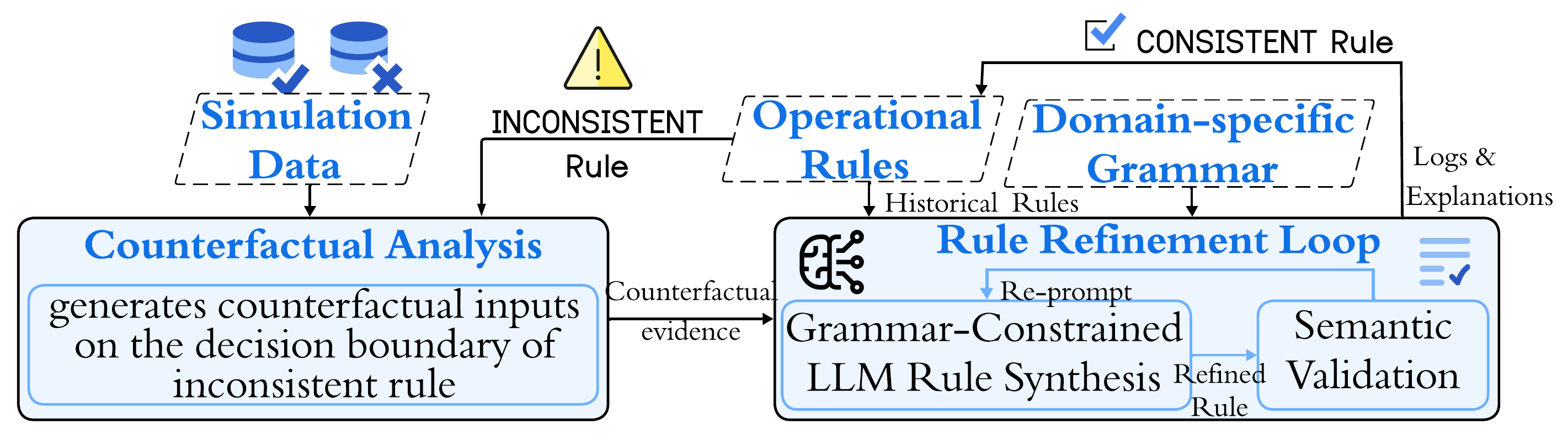}
    \caption{Operational Rule Refinement Approach Overview.}
    \Description{Overview of the operational rule refinement pipeline showing inputs (rule set, grammar, labeled executions), a counterfactual step to localize inconsistency boundaries, a grammar constrained LLM refinement loop, and validation producing an updated rule.}
    \label{fig:llm-framework}
\end{figure}

\section{Operational Rule Refinement Framework}
\label{sec:approach}
To address inconsistencies between operational rules and observed system behavior, we introduce a grammar-constrained rule refinement framework that leverages counterfactual reasoning combined with grammar-constrained LLM to produce interpretable, syntactically and semantically valid refinements of operational rules. The framework can be applied as a corrective mechanism for outdated operational rules within safety specifications that have become inconsistent with evolving system behavior.

As a running example, we consider an autonomous driving controller parameterized by three scenario features; the ego vehicle speed (\texttt{ego\_speed}), the longitudinal distance to the front vehicle (\texttt{dist\_front}), and the lateral offset within the lane (\texttt{lane\_offset}). In the context of autonomous driving, these features directly influence lane keeping behavior and collision risk. Operational rules encode admissible regions of the input space in which the unchanged safety requirement is intended to hold, i.e., minimum headway constraints at a given speed or allowable lane deviation under certain road conditions. Such rules capture the safety requirement into conditions over the variables, and may lose consistency with observed controller behavior as the operating environment changes or evolves. When these rules become inconsistent, they can yield misleading verdicts even if the underlying safety requirement is unchanged, motivating refinement of the rules rather than adaptation of the safety specification.

\subsection{Operational Rules and Semantics}

Let $X \subseteq \mathbb{R}^d$ denote the input domain of the system under test (SUT), where each input vector $x = [i_1, \ldots, i_d]$ represents input variables. The label of $x$ is $y$, where $y=\mathsf{Pass}$ if the observed system outcome satisfies the safety requirement, and $y=\mathsf{Fail}$ if it violates the safety requirement. We assume these outcome labels reflect the SUT simulator ground truth for the underlying safety requirement, handling label noise is outside our refinement objectives.

An \emph{operational rule} is a predicate $r : X \to \{\mathsf{true}, \mathsf{false}\}$ that constrains the operating region of the system. Here, $\mathsf{true}$ and $\mathsf{false}$ denote the logical evaluation of the rule on a given input. Specifically, $r(x) = \mathsf{true}$ means that the rule holds for $x$, i.e., the input is considered valid within the admissible operating region, while $r(x) = \mathsf{false}$ indicates that the rule does not hold. For example, given a rule $r_1:\ (\texttt{dist\_front} < 5.0)\ \wedge\ (\texttt{ego\_speed} > 0)$ and an input vector $x_1=\texttt{(ego\_speed=8.0,\ dist\_front=4.2,\ lane\_offset=0.1)}$, $r_1(x_1)=\mathsf{true}$, indicating that $x_1$ is valid and satisfies the admissible condition expressed by $r_1$. This indicates that, at speed $8.0$ the ego vehicle is operating in a safe state where the front vehicle is within $5.0$ units of distance, so the scenario lies inside the region of the input space that $r_1$ designates as admissible for the system's controller with respect to the safety requirement.

\begin{table}
\small
\centering
\caption{Consistency of an operational rule $r$ on a given input vector $x$ with respect to its rule set group and outcome $y$ of $x$.}
\Description{This table describes the interpretation of an operational rule on an input vector given the rule set to which the rule belongs and the observed system outcome. A rule set determines the rule verdict.}
\label{tab:outcome_groups}
\begin{tabular}{lllll}
\toprule
\textbf{Rule Set} &  \textbf{Rule Verdict} &\textbf{$r(x)$} & \textbf{Observed $y$} & \textbf{Consistency} \\
\midrule
\(R_{\mathsf{Pass}}\) & $\mathsf{Pass}$ & \(\mathsf{true}\) & \( \mathsf{Pass}\) & \emph{Consistent} \\
\(R_{\mathsf{Fail}}\) & $\mathsf{Fail}$&\(\mathsf{true}\) & \(\mathsf{Pass}\) & \emph{Inconsistent} \\
\(R_{\mathsf{Pass}}\) & $\mathsf{Pass}$&\(\mathsf{true}\) & \(\mathsf{Fail}\) & \emph{Inconsistent} \\
\(R_{\mathsf{Fail}}\) & $\mathsf{Fail}$&\(\mathsf{true}\) & \( \mathsf{Fail}\) & \emph{Consistent} \\
$R_{\mathsf{Pass}}$ & $\mathsf{Pass}$ &\(\mathsf{false}\) & -- & \emph{Inconclusive} \\
$R_{\mathsf{Fail}}$ & $\mathsf{Fail}$ &\(\mathsf{false}\) & -- & \emph{Inconclusive} \\
\bottomrule
\end{tabular}
\end{table}

We consider two rule sets, $R_{\mathsf{Pass}}$ and $R_{\mathsf{Fail}}$, encoding all operational rules $r$ that imply \emph{Pass} or \emph{Fail}, respectively, on a given input $x$. Table~\ref{tab:outcome_groups} summarizes how an operational rule $r$ is interpreted on an input vector $x$ given the rule set to which $r$ belongs and the observed system outcome $y$ for $x$.
A rule set determines the rule verdict. Rules in $R_{\mathsf{Pass}}$ assign a $\mathsf{Pass}$ verdict when they hold on $x$, while rules in $R_{\mathsf{Fail}}$ assign a $\mathsf{Fail}$  verdict when they hold on $x$.
The column $r(x)$ reports whether the rule holds on $x$. When $r(x)=\mathsf{true}$, the rule makes a definitive verdict and its consistency is determined by comparing that verdict to the observed outcome $y$. A rule in $R_{\mathsf{Pass}}$ is \emph{consistent} if $y=\mathsf{Pass}$, i.e., the rule holds on $x$ and correctly assigns a passing verdict, and \emph{inconsistent} if $y=\mathsf{Fail}$, i.e., the rule holds on $x$ but incorrectly assigns a $\mathsf{Pass}$ verdict while the observed outcome is a failure. 
A rule in $R_{\mathsf{Fail}}$ is \emph{consistent} if $y=\mathsf{Fail}$, i.e., the rule holds on $x$ and correctly assigns a failing verdict, and \emph{inconsistent} if $y=\mathsf{Pass}$, i.e., the rule holds on $x$ but incorrectly assigns a $\mathsf{Fail}$ verdict while the observed outcome is a pass. When $r(x)=\mathsf{false}$, the rule does not apply to $x$ and therefore yields no definitive verdict. This case is labeled \emph{inconclusive} and the outcome $y$ is not used (shown as ``--''). In this paper, we refine inconsistent rules, since they expose discrepancies between operational rules and observed system behavior. In our running example, consider $r_1 \in R_{\mathsf{Pass}}$ as a pass rule and the observed outcome of $x_1$ is $y_1=\textit{Fail}$, which makes $r_1$ inconsistent. This indicates that $x_1$ satisfies the admissible condition encoded by $r_1$, so the rule assigns a pass verdict, yet the system execution on $x_1$ violates the safety requirement.

\label{sec:grammar}
Each operational rule is expressed using a domain-specific grammar that defines 
the syntactic space of valid predicates over the system inputs. In general, the structure of the grammar is guided by domain knowledge, which determines the form and semantics of admissible expressions. In the automotive domain, operational rules for cyber-physical systems are typically expressed as arithmetic and relational constraints 
over configuration parameters and input variables. While it may vary across domains, in this paper we adopt the grammar $G$ introduced in prior work~\cite{gaaloul2021combining} 
to express environmental assumptions for cyber-physical systems, as it is particularly suited to the automotive context studied here. Each operational 
\noindent
\begin{minipage}[t]{0.5\columnwidth}
\fbox{%
\begin{minipage}[t]{0.96\linewidth}
\small
\[
\begin{aligned}
\text{Rule} &::= \text{Disj}\\
\text{Disj} &::= \text{Disj} \lor \text{Conj} \mid \text{Conj}\\
\text{Conj} &::= \text{Conj} \land \text{Rel} \mid \text{Rel}\\
\text{Rel}  &::= \text{Exp}\ \text{rop}\ \text{Exp}\\
\text{rop}  &::= < \mid \le \mid > \mid \ge \mid = \mid \neq\\
\text{Exp}  &::= \text{Exp}\ \text{aop}\ \text{Exp} \mid \text{const} \mid \text{var}\\
\text{aop}  &::= + \mid - \mid * \mid /
\end{aligned}
\]
\end{minipage}}
\end{minipage}\hfill
\begin{minipage}[t]{0.45\columnwidth}
rule is a hierarchical logical formula built from disjunctions ($Disj$), conjunctions ($Conj$), and relational predicates ($Rel$) over arithmetic expressions ($Exp$). A rule is composed of one or more disjunctive clauses, each representing an alternative valid operating condition. Each predicate $Rel$
\end{minipage}

 \noindent compares two expressions $Exp$ using operators $\{<,\le,>,\ge,=,\neq\}$, where expressions are formed from arithmetic operations over constants ($const$) and input variables ($var$).

\textbf{Problem Statement.} Given the rule sets $R_{\mathsf{Pass}}$ and $R_{\mathsf{Fail}}$, and a labeled input set 
$T = \{(x_i, y_i)\}$, the objective is to derive refined rule sets as changes in the components of existing rules, such as constants, input variables, relational operators, 
or logical connectors, so that they satisfy the following conditions: (i) Maintain semantic consistency with the grammar; (ii) Reduce the number of inconsistent rules; (iii) Preserve previously consistent rules; (iv) Eliminate contradictory rules. Note that the safety requirement and the stated ODD remain fixed. We refine operational rules that capture acceptance criteria for verification when their verdicts become inconsistent with observed outcomes. A refined rule is therefore a hypothesis about admissible operating conditions, not an authorization to broaden the safety envelope in response.

\subsection{Approach Overview} The grammar-constrained rule refinement framework operates on simulation or testing data and existing operational rules to resolve inconsistencies between operational rules and observed system behavior. In this work, we assume the presence of a safety operational rule set evaluated for inconsistencies using a manual review process or an automated consistency checking mechanism. We also assume that the rules are expressed in temporal logic according to a predefined grammar that constrains the admissible structure of rules (e.g., grammar $G$). Figure~\ref{fig:llm-framework} shows the overview of our approach. 
The framework takes as input (1) an operational rule set $\mathcal{R}$ containing at least one inconsistent rule $r$, (2) a grammar specification $G$ (3) a labeled execution dataset $\mathcal{D}$ of simulation or test cases. Each test case contains an input vector $x$ of input values and an observed outcome $y \in {\textit{Pass}, \textit{Fail}}$ with respect to a given safety specification. For example, the input vector $x_1$, $r_1$ assigns the $x_1$ a \textit{Pass} verdict. However, the observed system outcome for $x_1$ is $y_1=\textit{Fail}$, which makes $r_1$ inconsistent. The approach then proceeds through the following steps: 

\textit{1) Counterfactual Analysis:} The counterfactual analysis step generates a counterfactual input  $x'$ together with its observed outcome label  $y'$ obtained by re-executing the system. Here,  $y'$ denotes the system outcome for  $x'$. Intuitively,  $x'$ shows a minimally perturbed input whose observed outcome aligns with the rule assigned verdict, thereby localizing a nearby decision boundary for the inconsistent rule. The step takes as input an inconsistent operational rule $r$ and a labeled simulation dataset $\mathcal{D}$ containing test inputs and observed outcomes. It produces a counterfactual evidence file $E$ by generating counterfactual inputs for inputs that expose inconsistencies in $r$. Concretely, for each inconsistent case $(x,y)\in\mathcal{D}$, we search for a minimally perturbed input $x'$ such that the verdict flips, yielding evidence of a local decision boundary for the predicates in $r$. For our running example, the Counterfactual Analysis step generates a counterfactual input
$x_1'=\texttt{(ego\_speed=8.0, dist\_front=4.0, lane\_offset=0.1)}$
with label $y'=\textit{Pass}$, indicating that a small change to \texttt{dist\_front} is sufficient to flip the observed outcome and localize a nearby decision boundary for predicates in $r$. For each counterfactual, we calculate the feature-wise perturbation as $\Delta = x' - x$. In our example, the perturbation is $-0.2$.
The counterfactual $x'$ is obtained through an $L_1$ minimal-change search~\cite{wachter2017counterfactual} over the input space, which identifies the smallest modification that restores agreement between the rule verdict and the observed system behavior. Starting from $x$, the search incrementally expands the $L_1$ radius and evaluates modified feature assignments until it finds the first configuration $x'$ that flips the verdict. The resulting evidence file $E$ stores the dataset $\mathcal{D}$, the rule $r$, the paired inputs $(x,x')$ together with their labels $(y,y')$ and perturbation $\Delta$.

\textit{2)  Rule Refinement Loop:} This step takes the evidence $E$ together with the grammar specification $G$, and historical consistent rules used for the semantic validation. First, a grammar-constrained LLM acts as a synthesis assistant to propose a candidate refinement of $r$. Through zero-shot instruction prompting with a reference format exemplar and constraint-based guidance, the prompt provides the paired boundary inputs $(x,y)$ and $(x',y')$ and the corresponding $\Delta$ from $E$ and instructs the LLM to produce $r'$, a refinement of $r$, with minimal, grammar-compliant predicate changes such as threshold adjustments, operator replacements, or selective addition or removal of conjuncts and disjuncts. The template is provided below:
\noindent\fbox{%
\begin{minipage}{0.98\columnwidth}
\small
\textbf{Prompt template.}

\emph{Input:} $G$; inconsistent rule $r$; historical rules; evidence.

\emph{Task:} Return a refined rule $r'$ in the syntax of $G$ and a short explanation.

\emph{Loop:} If $r'$ uses out-of-vocabulary tokens or conflicts with historical rules, re-prompt with the failure summary and regenerate.

\emph{Format exemplar:} $(0<\textit{ARG2}<5)\wedge(\textit{ARG1}>0)\ \lor\ (8<\textit{ARG2}<12)$
\end{minipage}}

The objective of the refinement is to restore consistency with the observed system behavior while preserving the rule’s original semantics and interpretability and remaining within the grammar constraints. For our running example, the grammar-constrained refinement identifies the predicate in $r_1$ most responsible for the inconsistency. Since $r_1$ is a \textit{Pass} rule whose verdict on $x_1$ is \textit{Pass} while the observed outcome is \textit{Fail}, the refinement makes the rule more restrictive to exclude the failing region. The LLM proposes tightening the \texttt{dist\_front} condition and the candidate refinement yields $r_1^{\star}: (\texttt{dist\_front} < 4.1) \wedge(\texttt{ego\_speed} > 0)$. Tightening the threshold from $5.0$ to $4.1$ therefore shrinks the admissible region so that it excludes the failing neighborhood while preserving the original intent of the rule and keeping the refinement minimal.

In addition to the objective, the prompt instructs the LLM to propose a candidate refinement that remains consistent with the rule set $\mathcal{R}$. The rule set consistency is enforced by checking rule candidates on the labeled executions $\mathcal{D}$ before acceptance, subject to the following conditions:
\begin{itemize}
    \item (i) \emph{No contradictions:} there is no $(x,y)\in\mathcal{D}$ such that a pass rule and a fail rule both hold on $x$. We use SMT satisfiability~\cite{de2008z3} to check the satisfiability of opposite class rules and flag any potential overlap that can be treated as a contradiction.
    \item (ii) \emph{Preserved consistency:} for any $(x,y)\in\mathcal{D}$ where a historical rule in $\mathcal{R}$ was consistent, adding the candidate refinement does not make that rule inconsistent;
    \item (iii) \emph{Target inconsistency resolved:} the candidate reduces mismatches between the target rule verdict and the observed outcome label. 
\end{itemize}

 If the candidate violates the allowed vocabulary, or fails these checks, the loop re-initializes the refinement process. 
Before termination, each surviving candidate undergoes semantic validation on $\mathcal{D}$ to ensure that it restores consistency without introducing new inconsistencies. 
The refinement rule $r^{\star}$ is returned only if it passes validation; otherwise, the loop continues. The grammar conformance is enforced through prompt-level constraints, including a whitelist of allowed tokens and operators (e.g., $\{\wedge,\vee,<,>,\le,\ge\}$) and domain feature names. The full prompt and the examples are provided in our shared package~\cite{anonymous2026datasetrev}. The output of the framework is (i) the refinement rule $r^{\star}$ expressed in the syntax of the grammar, (ii) the change log summarizing the changes applied to $r$, and (iii) a short explanation of how the refinement addresses the observed inconsistencies.

\section{Experimental Exploration}
\label{sec:eval}
In this section, we conduct a first experimental exploration on our running case-study system, an autonomous driving control system~\cite{DBLP:conf/sbst/BiagiolaK24} that implements autopilot control for both lateral and longitudinal guidance in lane-following scenarios. The input vector includes numeric signals from both the ego-vehicle and the environment such as speed, steering angle, road curvature, weather, and obstacle distance. The system processes the inputs to compute throttle and steering adjustments that ensure lane keeping. The safety requirement is that \emph{the vehicle maintains its lane within admissible bounds}. We randomly generate 198 inputs, execute one run per input, and label each run \emph{Pass} if the requirement holds, and \emph{Fail} otherwise. To evaluate our framework, we report five evaluation metrics:

\emph{\textbf{(1) Decisiveness gain (DG)}} measures how consistently the rule's verdict matches the actual simulation outcomes. We compute $1 -\frac{N_{\mathit{mismatch}}}{N}$, where $\mathit{N_{mismatch}}$ denotes the number of runs for which the rule does not match the ground truth verdict, and total runs $N$. 

\textbf{(2) \emph{Semantic validity (SV)}} measures whether the refined rule stays grounded in the provided ODD and current operational rules. We use expert ratings to mark a predicate as \emph{invalid} if it introduces an out-of-range bound, a variable not present in the input vector, or an unsupported operator that violates the grammar or data constraints. We compute $1-\frac{N_{\mathit{invalid}}}{N_{\mathit{pred}}}$, where $N_{\mathit{invalid}}$ is the number of invalid predicates and $N_{\mathit{pred}}$ is the total number of predicates.

\textbf{(3) \emph{Interpretability (I)}} measures whether the LLM explanation is easy to follow and justifies all refinements. We use expert ratings: $1.0$ (Excellent) if the explanation (i) identifies and isolates inconsistencies in the original rule, (ii) presents the refined rule, and (iii) clearly justifies each major change. We assign $0.7$ to $0.8$ (High) when the explanation may not cover every edit but provides specific, well matched justifications for the main refinements. We assign $0.5$ (Low) when the explanation remains generic and does not justify the specific changes that were needed for most predicates.

\textbf{(4) \emph{Grammar compliance (GC)}} measures the structural correctness of the refined rule, i.e., whether it preserves the grammar’s disjunctive and conjunctive structure and follows the format exemplar in the prompt. We tokenize the rule (e.g., \texttt{`operator'}, \texttt{ARG}, \texttt{Value}, `(', `(') and count the structural violations as tokens that break the grammar. We then compute $1 - \frac{N_{\mathit{viol}}}{N_{\mathit{tok}}}$, where $N_{\mathit{viol}}$ is the number of violating tokens and $N_{\mathit{tok}}$ is the total number of tokens.

\textbf{(5) \emph{Change minimality (CM)}} measures how conservatively the LLM refines the rule while preserving the original constraints. We use expert ratings: $1.0$ (Optimal) for pruning to the logical core with minimal edits and the same variables (e.g., \texttt{ARG2 > 3 AND ARG2 > 5} $\rightarrow$ \texttt{ARG2 > 5}). We assign $0.7$ to $0.8$ (Conservative) for moderate cleanup or added complementary operators (e.g., adding an upper bound to \texttt{ARG2 > 5}) without changing variables. We assign $0.4$ to $0.5$ (Over constrained) when the rule adds unjustified bounds that narrow its scope (e.g., \texttt{ARG1 > 0} $\rightarrow$ \texttt{1 < ARG1 < 2}). We assign $0.0$ to $0.3$ (Low) for extensive rewrites where most predicates change and the rule logic shifts substantially.

\noindent\textbf{Conventional baseline.} We adopt a genetic programming based baseline from prior work for inferring grammar constrained assertions~\cite{gaaloul2021combining,jodat2025automated} that assign pass and fail verdicts to system test inputs. Prior results report high inconsistency rates, quantified using accuracy and misprediction metrics across multiple systems and requirements. From the prior work, we used a dataset of $N=198$ labeled runs and a representative pass rule for an ADS controller system in the open-source BeamNG simulator~\cite{beamng}. We observed a decisiveness gain of $\mathit{DG}=0.86$, with $N_{\mathit{mismatch}}=27$ runs, indicating that the pass rule was inconsistent. We then applied our approach to this inconsistent rule using $8$ LLM variants~\cite{zhao2023survey}: GPT-5 (Thinking and Instant), Claude Sonnet 4.5, DeepSeek (DeepThinking and Normal), Qwen3 Max, and Gemini 1.5 (Pro and Flash). Under the same grammar $G$ and prompt template, we recorded the refined rules along with a change log and explanation. We then computed decisiveness on the $198$ labeled runs for all variants. All variants reduced mismatches to $0$ ($\mathit{DG}=1.0$), corresponding to a gain of $+0.14$ over the baseline. These results indicate that our approach resolved the inconsistencies of the original rule and improved alignment between operational rule verdicts and observed system behavior.

\textbf{LLM Variant Study.} Given the promising preliminary observations, we assess how refinement quality varies with model choice in our language-in-the-loop setting and we retrieve lessons learned about LLM use in safety operational contexts. We analyze the refined rules, change logs, and explanations generated by our approach configured with the $8$ LLM variants.

\begin{table}[t]
\centering
\caption{Evaluation metric scores per LLM variant.}
\Description{This table reports, for each evaluated LLM variant, the scores of four evaluation metrics: GC (grammar compliance), SV (semantic validity), I (interpretability), and CM (change minimality). Higher values indicate better performance on the corresponding metric; boldface highlights the best observed scores.}
\label{tab:llm_metrics}
\renewcommand{\arraystretch}{1}
\small
\resizebox{\linewidth}{!}{%
\begin{tabular}{lcccc|lcccc}\toprule
LLM&GC&SV&I&CM&LLM&GC&SV&I&CM\\\midrule
GPT5 Thinking&\textbf{1.0}&\textbf{1.0}&0.5&\textbf{0.9}&Qwen3 Max&\textbf{1.0}&0.8&0.7&0.7\\
GPT5 Instant&\textbf{1.0}&0.7&0.7&0.4&DeepSeek DeepThinking&0.5&0.5&0.7&0.3\\
Gemini Flash 2.5&\textbf{1.0}&\textbf{1.0}&0.8&0.7&DeepSeek Normal&0.7&\textbf{1.0}&0.5&0.8\\
Gemini Pro 2.5&\textbf{1.0}&0.4&0.7&0.2&Claude Sonnet 4.5&\textbf{1.0}&0.2&\textbf{1.0}&0.0\\
\bottomrule\end{tabular}}
\end{table}

Table~\ref{tab:llm_metrics} reports four metric scores for each LLM. GPT5 Thinking mode shows the strongest combination of semantic validity and minimality while staying fully grammar compliant. Gemini Flash 2.5 and Qwen3 also remain grammar compliant with relatively strong semantic validity. In contrast, Gemini Pro 2.5 and Claude Sonnet 4.5 are grammar compliant and fairly interpretable. DeepSeek exhibits mixed behavior, with the Normal variant achieving high semantic validity and minimality but lower interpretability, while DeepThinking shows weaker grammar compliance and more extensive changes. We retrieve the following lessons:

\textbf{Lesson~1:} Even with the same grammar, prompt, and simulation data, different LLMs vary in outcome. Some rules look correct but include formatting that breaks the expected structure and variable naming. These should be treated as unsafe to apply. For example, DeepSeek DeepThinking returned the rule inside a markdown code block and wrapped the rule with an extra outer list , which violates the structure. The expected rule is $\texttt{[('greater\_than\_func','ARG1','0')]}...$, but the model returned $\texttt{[[('greater\_than\_func','ARG1','0')]...]}$.

\textbf{Lesson~2:} LLMs tend to increase apparent safety by tightening bounds and occasionally adding extra constraints, but this can over constrain the rule in a conservative way that is not correctly grounded in the provided ODD, yielding many unnecessary nominal restrictions. For example, a refined rule may turn a simple threshold into a tight range, changing \texttt{0 < ARG1} to \texttt{0 < ARG1 < 8}. This can look safer, yet it may be unsupported by the provided ODD and therefore unjustified. In safety critical use, validity checks against the ODD should be adopted to flag new or tightened bounds that are not semantically valid, and an iterative feedback loop should be triggered whenever the model makes large threshold shifts, violates the grammar, or introduces new variables, even if the explanation appears convincing.

\textbf{Lesson~3:} There is a link between how much a model changes the rule and how easy its output is to interpret. When changes are broader,  the model explicitly critiques the original inconsistencies and justifies each major change. For example GPT5 Thinking mode makes small, targeted refinements, reducing the rule to \texttt{ARG1 > 0} or \texttt{ARG2 > 3}, and its justification only focuses on redundancy removal and fixing a malformed predicate. In contrast, Gemini Pro 2.5 introduces new conjunctive constraints and new variables (\texttt{ARG3}) alongside operator changes, and it provides a structured rationale for each addition, for example explaining the shift to \texttt{>=} at the boundary and motivating the new \texttt{<} caps. 

This initial study shows that grammar-constrained, counterfactual-guided refinement can eliminate baseline inconsistencies, while the LLM variant study reveals model dependent quality and safety trade-offs that illustrate \textit{``what can go wrong''}. Although grammar guidance supports syntactic correctness, safety guarantees require additional strategies. Additional limitations and future directions are summarized in the next section.

\section{Conclusion and Future Work}
\label{sec:conclusion}
Safety operational rules can lose alignment with observed system behavior as systems and operating environments evolve. 
This paper introduced a rule refinement framework that combines counterfactual reasoning with a grammar-constrained LLM refinement loop to produce interpretable refinements that are syntactically correct and semantically valid. 
An initial study on an autonomous driving subsystem showed that our loop eliminates inconsistencies produced by the selected conventional method, with $+0.14$ decisiveness. 
An LLM variant study further exposed model dependent quality and safety trade offs, including syntactic violations and overly conservative refinements that risk overfitting to the observed dataset. 
\textbf{Limitations and Future Work.} The results of our initial exploration showed that grammar guidance alone does not provide safety guarantees and may still yield semantically unjustified refinements under limited evidence. Our current study is preliminary and focuses on a single subsystem and dataset. Accordingly, future work focuses on (i) strengthening grammar enforcement via a strongly typed rule generator and a parser based acceptance mechanism that rejects any output violating the grammar or structural format; 
(ii) reinforcing semantic validation beyond a static regression test suite through simulation based falsification and robustness testing to ensure consistency over a broader input space and the stated ODD, while flagging unjustified tightened constraints and exposing unsafe overfitting; and 
(iii) mitigating overly conservative refinements by incorporating change minimality into an independent selection mechanism so edits are penalized even when decisiveness is high. 
For the evaluation, we will broaden baselines and study subjects to determine whether grammar-constrained LLM refinement offers clear benefits over established interpretable rule learning and specification mining, including decision trees and decision rules and temporal specification mining methods. 
We will then scale experiments across multiple ADS subsystems, requirements, and ODDs, and study how expanded grammars, prompting strategies, and robust validation mechanisms affect refinement quality and reduce safety risks.

\begin{acks}
This material is based upon work supported by the National Science Foundation under Grant No. 2347294. Any opinions, findings, and conclusions or recommendations expressed in this material are those of the author(s) and do not necessarily reflect the views of the National Science Foundation.
\end{acks}

\bibliographystyle{ACM-Reference-Format}
\bibliography{bibliography}

@book{fraade2018measuring,
  title        = {Measuring Automated Vehicle Safety: Forging a Framework},
  author       = {Fraade-Blanar, Laura and Blumenthal, Marjory S. and Anderson, James M. and Kalra, Nidhi},
  year         = {2018},
  publisher    = {RAND Corporation},
  address      = {Santa Monica, CA},
  isbn         = {978-1-9774-0164-9},
  url          = {https://www.rand.org/pubs/research_reports/RR2662.html}
}

@article{garcia2022vehicle,
  title={From the vehicle-based concept of operational design domain to the road-based concept of operational road section},
  author={Garc{\'\i}a, Alfredo and Llopis-Castell{\'o}, David and Camacho-Torregrosa, Francisco Javier},
  journal={Frontiers in Built Environment},
  volume={8},
  pages={901840},
  year={2022},
  publisher={Frontiers Media SA}
}

@article{debouk2019overview,
  title={Overview of the second edition of ISO 26262: Functional safety—Road vehicles},
  author={Debouk, Rami},
  journal={Journal of System Safety},
  volume={55},
  number={1},
  pages={13--21},
  year={2019}
}

@inproceedings{kirovskii2019driver,
  title={Driver assistance systems: analysis, tests and the safety case. ISO 26262 and ISO PAS 21448},
  author={Kirovskii, OM and Gorelov, VA},
  booktitle={IOP Conference Series: Materials Science and Engineering},
  volume={534},
  number={1},
  pages={012019},
  year={2019},
  organization={IOP Publishing}
}

@inproceedings{tuncali2018simulation,
  title={Simulation-based adversarial test generation for autonomous vehicles with machine learning components},
  author={Tuncali, Cumhur Erkan and Fainekos, Georgios and Ito, Hisahiro and Kapinski, James},
  booktitle={2018 IEEE intelligent vehicles symposium (IV)},
  pages={1555--1562},
  year={2018},
  organization={IEEE}
}

@article{zhao2023survey,
  title={A survey of large language models},
  author={Zhao, Wayne Xin and Zhou, Kun and Li, Junyi and Tang, Tianyi and Wang, Xiaolei and Hou, Yupeng and Min, Yingqian and Zhang, Beichen and Zhang, Junjie and Dong, Zican and others},
  journal={arXiv preprint arXiv:2303.18223},
  volume={1},
  number={2},
  year={2023}
}

@inproceedings{pnueli1977temporal,
  title={The temporal logic of programs},
  author={Pnueli, Amir},
  booktitle={18th annual symposium on foundations of computer science (sfcs 1977)},
  pages={46--57},
  year={1977},
  organization={ieee}
}

@book{rescher2012temporal,
  title={Temporal logic},
  author={Rescher, Nicholas and Urquhart, Alasdair},
  volume={3},
  year={2012},
  publisher={Springer Science \& Business Media}
}

@inproceedings{asarin2011parametric,
  title={Parametric identification of temporal properties},
  author={Asarin, Eugene and Donz{\'e}, Alexandre and Maler, Oded and Nickovic, Dejan},
  booktitle={International Conference on Runtime Verification},
  pages={147--160},
  year={2011},
  organization={Springer}
}

@article{hoxha2018mining,
  title={Mining parametric temporal logic properties in model-based design for cyber-physical systems},
  author={Hoxha, Bardh and Dokhanchi, Adel and Fainekos, Georgios},
  journal={International Journal on Software Tools for Technology Transfer},
  volume={20},
  number={1},
  pages={79--93},
  year={2018},
  publisher={Springer}
}

@article{kyriakis2019specification,
  title={Specification mining and robust design under uncertainty: A stochastic temporal logic approach},
  author={Kyriakis, Panagiotis and Deshmukh, Jyotirmoy V and Bogdan, Paul},
  journal={ACM Transactions on Embedded Computing Systems (TECS)},
  volume={18},
  number={5s},
  pages={1--21},
  year={2019},
  publisher={ACM New York, NY, USA}
}

@article{gaaloul2021combining,
  title={Combining genetic programming and model checking to generate environment assumptions},
  author={Gaaloul, Khouloud and Menghi, Claudio and Nejati, Shiva and Briand, Lionel C and Parache, Yago Isasi},
  journal={IEEE Transactions on Software Engineering},
  volume={48},
  number={9},
  pages={3664--3685},
  year={2021},
  publisher={IEEE}
}

@inproceedings{DBLP:conf/sbst/BiagiolaK24,
  author       = {Matteo Biagiola and
                  Stefan Klikovits},
  title        = {{SBFT} Tool Competition 2024 - Cyber-Physical Systems Track},
  booktitle    = {Proceedings of the 17th {ACM/IEEE} International Workshop on Search-Based
                  and Fuzz Testing, {SBFT} 2024, Lisbon, Portugal, 14 April 2024},
  pages        = {33--36},
  publisher    = {{ACM}},
  year         = {2024},
  url          = {https://doi.org/10.1145/3643659.3643932},
  doi          = {10.1145/3643659.3643932},
  bibsource    = {dblp computer science bibliography, https://dblp.org}
}

@misc{anonymous2026datasetrev,
  author       = {{Khouloud Gaaloul, Zaid Ghazal, Madhu Latha Pulimi, Sam Emmanuel Kathiravan}},
  year         = {2026},
  title        = {Additional Materials},
  howpublished = {\url{https://replication66.github.io/SEAMS2026/}}
}

@article{wachter2017counterfactual,
  title={Counterfactual explanations without opening the black box: Automated decisions and the GDPR},
  author={Wachter, Sandra and Mittelstadt, Brent and Russell, Chris},
  journal={Harv. JL \& Tech.},
  volume={31},
  pages={841},
  year={2017},
  publisher={HeinOnline}
}

@article{jodat2025automated,
  title={Automated Test Oracles for Flaky Cyber-Physical System Simulators: Approach and Evaluation},
  author={Jodat, Baharin A and Gaaloul, Khouloud and Sabetzadeh, Mehrdad and Nejati, Shiva},
  journal={arXiv preprint arXiv:2508.20902},
  year={2025}
}

@inproceedings{nouri2024engineering,
  title={Engineering safety requirements for autonomous driving with large language models},
  author={Nouri, Ali and Cabrero-Daniel, Beatriz and T{\"o}rner, Fredrik and Sivencrona, H{\aa}kan and Berger, Christian},
  booktitle={2024 IEEE 32nd International Requirements Engineering Conference (RE)},
  pages={218--228},
  year={2024},
  organization={IEEE}
}

@article{li2025automatic,
  title={Automatic Generation of Safety-compliant Linear Temporal Logic via Large Language Model: A Self-supervised Framework},
  author={Li, Junle and Tian, Meiqi and Zhong, Bingzhuo},
  journal={arXiv preprint arXiv:2503.15840},
  year={2025}
}

@article{jha2019telex,
  title={TeLEx: learning signal temporal logic from positive examples using tightness metric},
  author={Jha, Susmit and Tiwari, Ashish and Seshia, Sanjit A and Sahai, Tuhin and Shankar, Natarajan},
  journal={Formal Methods in System Design},
  volume={54},
  number={3},
  pages={364--387},
  year={2019},
  publisher={Springer}
}

@inproceedings{lemieux2015general,
  title={General LTL specification mining (T)},
  author={Lemieux, Caroline and Park, Dennis and Beschastnikh, Ivan},
  booktitle={2015 30th IEEE/ACM International Conference on Automated Software Engineering (ASE)},
  pages={81--92},
  year={2015},
  organization={IEEE}
}

@article{jodat2024test,
  title={Test generation strategies for building failure models and explaining spurious failures},
  author={Jodat, Baharin A and Chandar, Abhishek and Nejati, Shiva and Sabetzadeh, Mehrdad},
  journal={ACM Transactions on Software Engineering and Methodology},
  volume={33},
  number={4},
  pages={1--32},
  year={2024},
  publisher={ACM New York, NY}
}

@article{zhu2022fuzzy,
  title={Fuzzy rule-based local surrogate models for black-box model explanation},
  author={Zhu, Xiubin and Wang, Dan and Pedrycz, Witold and Li, Zhiwu},
  journal={IEEE Transactions on Fuzzy Systems},
  volume={31},
  number={6},
  pages={2056--2064},
  year={2022},
  publisher={IEEE}
}

@inproceedings{de2008z3,
  title={Z3: An efficient SMT solver},
  author={De Moura, Leonardo and Bj{\o}rner, Nikolaj},
  booktitle={International conference on Tools and Algorithms for the Construction and Analysis of Systems},
  pages={337--340},
  year={2008},
  organization={Springer}
}

@ONLINE{beamng,
title = {BeamNG.tech},
url = {https://beamng.tech},
year = {Accessed: January 2026}
}

\end{document}